**A reassessment of the “hard-steps” model for the evolution of intelligent life**

**Short title: An alternative model of biospheric evolution**

Daniel B. Mills[1,2,3*], Jennifer L. Macalady[2,3,4,5], Adam Frank[6], Jason T. Wright[2,3,7]

[1]Department of Earth and Environmental Sciences, Paleontology & Geobiology, Ludwig-Maximilians-Universität München, 80333 Munich, Germany
[2]The Penn State Extraterrestrial Intelligence Center, Penn State, University Park, PA 16802 USA
[3]Center for Exoplanets and Habitable Worlds, Penn State, University Park, PA 16802 USA
[4]Department of Geosciences, Penn State, University Park, PA 16802 USA
[5]Astrobiology Research Center, Penn State, University Park, PA 16802 USA
[6]Department of Physics and Astronomy, University of Rochester, Rochester, NY 14620, USA
[7]Department of Astronomy and Astrophysics, Penn State, University Park, PA 16802 USA

*Corresponding author (d.mills@lmu.de)

**Teaser/One-sentence summary:** Humans – and analogous life beyond Earth – may represent the probable outcome of biological and planetary co-evolution

**Abstract**
According to the “hard-steps” model, the origin of humanity required “successful passage through a number of intermediate steps” (so-called “hard steps”) that were intrinsically improbable in the time available for biological evolution on Earth. This model similarly predicts that technological life analogous to human life on Earth is “exceedingly rare” in the Universe. Here, we critically reevaluate core assumptions of the hard-steps model through the lens of historical geobiology. Specifically, we propose an alternative model where there are no hard steps, and evolutionary singularities required for human origins can be explained via mechanisms outside of intrinsic improbability. Furthermore, if Earth’s surface environment was initially inhospitable not only to human life, but also to certain key intermediate steps required for human existence, then the timing of human origins was controlled by the sequential opening of new global environmental windows of habitability over Earth history.

**INTRODUCTION – What is the “hard-steps” model?**

In 1983, the physicist Brandon Carter concluded that the time it took for humans to evolve on Earth (relative to the total lifespan of the Sun) suggests that our evolutionary origin was intrinsically unlikely, and that comparable human-like observers beyond the Earth are rare (*1*). Carter arrived at this conclusion by noting the order-of-magnitude coincidence (within a factor of about two) between the age of the Earth as it now appears to us ($t_e \approx 0.5 \times 10^{10}$ years, roughly equivalent to the timing of our emergence) and the estimated main sequence lifespan of the Sun ($\tau_0 = 10^{10}$ years, corresponding roughly to the habitable lifetime of the Earth). If one assumes, as Carter did, that the biological processes dictating evolutionary timescales on Earth and the physical processes determining the main sequence lifetime of the Sun "have nothing directly to do with each other" (*1*), then there is no *a priori* reason for predicting such a close "observational coincidence" between these two timescales (*2*). Carter, noting that evolutionary theory is unable to predict the "expected average time" ($t_i$) intrinsically required to evolve "intelligent observers," evaluated the different possibilities relating these three timescales to broadly constrain $t_i$: (i) $t_i << \tau_0$; (ii) $t_i \approx t_e$; and (iii) $t_i >> \tau_0$. Carter rejected (i) on the grounds that if it were true, then $t_e$ should have a much smaller value than $\tau_0$ (we should find ourselves on a much younger Earth), and "it is hard to think of any particular reason why our arrival should have been greatly delayed relative to the intrinsically expected time [$t_i$]" (*1*). Next, Carter rejected (ii) as "much less plausible *a priori*" than the alternatives, and recommended considering it only if convincing *a posteriori* evidence against the two remaining possibilities were to arise (*1*). Carter ultimately settled on (iii), arguing that if one accepts $t_i >> \tau_0$, then $t_e \approx \tau_0$ becomes explicable – indeed expected – by applying the self-selection (or "anthropic") principle (*3*), in that if we are going to evolve, we must necessarily evolve prior to $\tau_0$ (*2*), and on timescales most probably approaching $\tau_0$. It is through the application of this anthropic reasoning that Carter predicted that our evolutionary origin was inherently improbable within the externally allotted time ($\tau_0$) (*4*), with the corollary that analogous "intelligent observers" beyond the Earth would be equally improbable.

To explain the inherent unlikelihood of human origins, Carter proposed that the evolutionary emergence of humans must have depended on the "successful passage through a number of intermediate steps" in which traits necessary for human existence were gained (*1*). If the mean time required for such an essential step is "small" relative to $\tau_0$, then Carter considered the step "easy" (*2*), happening "effectively deterministically" (*2*) with "virtual certainty" in the provided time ($\tau_0$) (*1*). However, if the intrinsic mean time required for an essential intermediate step is "large" compared to $\tau_0$ (*2*) – "at least a significant fraction" of $\tau_0$ (*1*) – then Carter variously considered such a step "critical" (*1*, *5*), "difficult" (*2*), or "hard" (*6*). Since "easy" steps are more or less guaranteed to occur with respect to $\tau_0$, it is only the "hard" steps that need to be considered in estimating the likelihood of human existence within $\tau_0$. Recognizing that the conditional probability ($P$) of successfully completing a number of equally unlikely steps ($n$) within time ($t$) follows the power law expression $P \approx t^n$, Carter initially proposed only one or two hard steps (*1*, *2*), as $n \geq 3$ predicts that steps would most likely be completed "very near" the upper limit of $\tau_0$ and that while our sun is "no longer young", the time remaining in its main sequence lifetime is nevertheless too great to reconcile with more than two steps (*2*). Favoring this logic, Carter ultimately concluded that at most one or two of the essential steps in our evolutionary history were truly "hard" within the bounds of the Sun's lifespan, and that the evolution of comparable biological beings on worlds beyond the Earth would similarly depend on "chance events with characteristic timescales long compared with those of stellar evolution" (*4*). With this formalization, the "hard-steps" model was born.

Since 1983, the hard-steps (or critical-steps or Carter) model has been utilized and refined by numerous authors (*7–14*), and remains a pervasive and influential framework for predicting the distribution and complexity of life beyond Earth (*15–19*). The hard-steps model also inspired the related concept of the "Great Filter" (Box 1), which more explicitly accounts for the claimed lack of evidence surrounding extraterrestrial intelligence (*20*). Consistent with Carter's predictions, proponents of the hard-steps model generally reiterate that human beings were an unlikely product of biological evolution on Earth, and that human-like life elsewhere in the observable universe is exceedingly rare. However, various aspects of the hard-steps model have been criticized, particularly the assumption that biospheric evolution unfolds independently of solar evolution (*8*, *21*, *22*), as well as the rejection of factors (for example, environmental) that could have conceivably "delayed" our evolutionary emergence in the scenario $t_i \ll \tau_0$ (*21*, *23*, *24*). Curiously, while these issues concern the evolutionary history of Earth's biosphere, comparatively few Earth historians (*12*, *14*, *16*, *21*), and even fewer evolutionary biologists (*25*), have responded to Carter's arguments in the literature, having primarily left astrophysicists, economists, and futurists to champion the hard-steps model unchecked. To correct for this historical imbalance, greater attention from the Earth and life sciences is needed to more exhaustively evaluate and test the core tenets of the hard-steps model.

In this paper, we challenge key fundamental assumptions of the hard-steps model through the lens of historical geobiology (*26*), the study of how Earth's surface environment and life have co-evolved over geologic time. In short, if Earth's surface environment was initially inhospitable not only to human life, but also to certain key intermediate steps in human evolution (e.g., the origin of eukaryotic cells, the origin of animal multicellularity), then the "delay" in our evolutionary origin can be best explained by the geological timescales necessary for creating the global-environmental conditions required for humans or human-analogues. In this view, we find ourselves so close to the upper limit of Earth's habitability because this is where the geologically narrow "window of human habitability" is located relative to Earth's total habitable lifespan.

For clarity, while we examine many of the published interpretations of the hard-steps model, as well as many of the published hard-steps candidates, our critique is more fundamental than this. We examine and challenge the original assumptions used to justify the hard-steps model in the first place, questioning whether the model is even necessary for explaining the temporal coincidence between human origins ($t_e$) and the predicted extinction of Earth's biosphere ($\tau_0$). Furthermore, we do not claim to have successfully falsified the hard-steps model, but we do claim to have formulated a justifiable and testable alternative to it. While our alternative model proposes that the evolutionary origin of humans or human-analogues was more predictable or probable than the hard-steps model concludes, we do not claim that the evolution of *Homo sapiens* in particular was "inevitable."

### What are the hard steps?

According to Carter, a hard (or critical) step must be both 1) essential to the evolutionary emergence of any given trait or organism, and 2) improbable with respect to the externally allotted time (*1*, *2*). Throughout his publications, Carter variously and interchangeably referred to the "emergence of civilization" (*1*), "the emergence of intelligent observers such as ourselves" (*2*), and "the evolution of what we recognize as intelligent life" (*4*) as the evolutionary innovation whose probability was ultimately in question. While the origin of a "scientific civilization such as our own" (*1*) and "the emergence of intelligent life" (*4*) more generally represent distinct evolutionary events in the history of life on Earth (that is, the origin of human civilization vs. the origin of *Homo sapiens*, respectively), the hard-steps model can be applied equally – as Carter phrased it – to any given "stage of advancement" (*1*), including "less 'advanced' stages of development" (*2*). While this language, betraying non-Darwinian notions of evolution as a linear "ascent" from lower to higher degrees of "advancement" (*27*), is perhaps too reminiscent of the Great Chain of Being for most modern evolutionary biologists to accept (*28*), Carter nevertheless rejected so-called "progressive" notions of evolution (with humans at the top) as "unduly anthropocentric" (*1*). Indeed, like the anthropic principle itself, the hard-steps model is applicable to humans and non-human entities alike – notably, extraterrestrial organisms (*1*), as well as any organism that has existed or will exist on Earth. In the various applications of the hard-steps model by other authors, the focus has primarily been on humans or *H. sapiens* (*7*, *11*), "intelligent life" (*9*, *18*), and "observerhood" (*12*). Others adopted a more operational approach relevant to the field of SETI (the search for extraterrestrial intelligence), defining intelligence as "the building of radio telescopes" (*8*) or life capable of manufacturing detectable "technosignatures" (*17*). In general, "intelligence" has no standard definition, and arguably agency (the capacity to deliberately change one's environment) and cognition (knowing how to perform these changes and reflecting on them) are more relevant traits for SETI (*29*). For our purposes, we are concerned with the existence of evolutionary transitions and processes (so-called "steps") that were both improbable (relative to $\tau_0$) and essential to the evolutionary origin of *H. sapiens* on Earth (since we are the self-reflective observers communicating about our own observations). While this effort concerns SETI, it applies equally to understanding evolutionary timescales on Earth (*30*), as well as how life in general may unfold on worlds beyond Earth.

Carter, originally envisioning only one or two "hard steps," initially proposed 1) the origin of the genetic code, and 2) what he phrased as "the final breakthrough in cerebral development" (*1*), presumably referring to a shift in cognition and behavior along the human lineage sometime after its split from its sister lineage, the Panina (today represented by chimpanzees and bonobos) (*31*) (Table 1). The physicists John Barrow and Frank Tipler provided the next set of candidate hard steps (totaling 10 overall; Table 1) in their 1986 book *The Anthropic Cosmological Principle*, along with three proposed criteria for identifying potential hard steps. Their first criterion was that "the step must have been unique; it must have occurred only once in the entire history of life" (*7*). In other words, hard steps must be what biologists call evolutionary "singularities," defined by the cell biologist Christian de Duve as "events or properties that have the quality of singleness, uniqueness" (*32*). Indeed, evolutionary biologists generally interpret traits that evolved only once in the history of life as improbable, reflecting the inherent contingency and unpredictability of the evolutionary process, and argue that such traits are unlikely to evolve again (*33–37*). This pattern contrasts with traits that were acquired convergently in numerous lineages, such as image-forming eyes, which evolved perhaps 40 different times (*38*), leading the evolutionary biologist Ernst Mayr to conclude that evolving eyes is "not at all improbable" (*35*). Barrow & Tipler cautioned, however, that a trait may appear unique not because it is improbable, but because numerous lineages acquired it and have since gone extinct with the exception of one. To avoid making this false positive, Barrow & Tipler's second criterion for the identification of hard steps was that the trait in question must be "polygenic" (coded by multiple genes), arguing that such traits are unlikely to disappear in a lineage without leaving traces in its descendants (*39*). This reasoning implies that if a polygenic trait appears to be unique across the tree of life, it is unique because it is inherently unlikely to evolve, not because it is the last of its kind (more on this reasoning below). Lastly, and perhaps redundantly, Barrow & Tipler's third criterion was that a hard step must be "essential for the existence of an intelligent species." Barrow & Tipler then justified their 10 proposed hard steps by virtue of these three criteria.

In the decades following Barrow & Tipler's 1986 book, astrophysicists, planetary scientists, mathematicians, futurists, and Earth system scientists proposed their own potential hard steps with varying degrees of overlap and estimates for $n$ (Table 1). Notably, compared to Carter's original estimate of $1 \leq n \leq 2$ (*1*), all subsequent efforts estimated $n > 2$ (Table 1). While Carter assumed Earth's biosphere will remain viable until the Sun leaves the main sequence (~5 billion years from now), subsequent Earth system modeling suggests the biosphere, or at least vascular land plant ecosystems, will go extinct – via $CO_2$ starvation and/or catastrophic warming – somewhere on the order of 1.0-1.9 billion years from now (*40–43*). In this latter scenario, $t_e$ goes from representing approximately ~50% of the externally allotted time for biological evolution to ~70-80%. This conclusion is consistent with a greater number of hard steps than Carter originally suggested (*6*, *12*), predicting an origin of humanity (assuming humans or human analogues evolve at all) closer to the outer edge of the externally allotted time.

With respect to the candidates themselves, the most popular suggestions (of those we surveyed) include the origin of 1) life ("abiogenesis"), 2) oxygenic photosynthesis, 3) eukaryotic cells ("eukaryogenesis"), 4) animal multicellularity, and 5) *H. sapiens* – the exact sequence proposed by Lingam & Loeb (*17*, *44*). For our purposes, these five evolutionary origins will henceforth serve as our working list of "candidate hard steps." Each of these candidates meets Barrow & Tipler's criteria in that they are 1) purported singularities, 2) coded by several genes, and 3) were essential to human existence (which is redundant for candidate #5). Note, while multicellularity in general has evolved numerous times (*45*, *46*), leading some to reject animal multicellularity as a candidate hard step (*14*, *18*), animal multicellularity has been considered an evolutionary singularity on cytological and ecological grounds (more on this below) (*47–49*). Furthermore, while the origin of *H. sapiens* is universally considered a hard step in our survey, there are anthropological grounds for questioning human uniqueness (Box 2). It must be noted that our five hard-step candidates are cumulative – that is, (5) could have only happened after (4), which could have only happened after (3), and so on. Next, only three of these five candidates are major transitions in evolution (MTE) (Box 3), with the origin of oxygenic photosynthesis and the origin of humans as the exceptions (*50*), and all occurred in the direct ancestors of humans, with the exception of the origin of oxygenic photosynthesis. Indeed, Barrow & Tipler emphasized that hard steps can fall outside the evolutionary lineage leading directly to humanity (*7*). Interestingly, other less popular candidate hard steps include more biospheric transitions, such as the permanent oxygenation of the atmosphere (*8*, *13*), or the colonization of the continents by plants, animals, and fungi (*8*), raising the question of whether global environmental and ecological transitions can qualify as hard steps (more in the following sections) (*12*).

To recap our criteria, Carter defined hard steps as both 1) improbable (within the externally allotted time), and 2) essential for the evolution of humanity (*1*). Barrow & Tipler added the criterion that hard steps should be singular events in the history of life on Earth (and they proposed a way to check for this singular status) (*7*). In addition to these criteria, Hanson (*9*) and Watson (*12*) subsequently proposed that hard steps should be spaced (roughly) evenly throughout the history of life. Therefore, candidate hard steps also need to be tested with respect to their timing in Earth history (*14*). Looking at our working list of five hard-step candidates (having already met Barrow & Tipler's criteria), the duration between steps averages 0.84 billion years (Gyr) with a standard deviation of 0.2-0.4 Gyr, depending on the intervals used, with the shortest temporal gaps ranging 0.5-0.6 Gyr and the longest gaps ranging 1.1-1.4 Gyr, thereby differing by only a factor of two to three (Fig. 1). Together, looking at the history of life on Earth, and applying the published criteria for identifying hard steps, it does appear that singular, evenly spaced transitions required for human existence occurred over Earth history, consistent with the existence of hard steps, and at least partially explaining the enduring popularity of the hard-steps model (*17*, *18*). Indeed, the hard-steps model is attractive because it offers an elegant and simple solution – by virtue of the self-selection principle – to why we observe ourselves so close in time to the predicted extinction of the biosphere. The roughly equal distribution of our hard-steps candidates in Earth history (Fig. 1) further demonstrates the apparent predictive power of the hard-steps model, reasonably bolstering its plausibility.

While the evidence outlined above arguably reinforces the existence of hard steps, the interpretation that these candidates were intrinsically improbable is primarily tied to their singular status. In other words, if these events were truly improbable relative to the lifespan of the biosphere, then the hard-steps model would be corroborated by our understanding of the history of life on Earth. However, if the singular nature of these candidates is questionable, or can be explained without resorting to improbability, then support for the hard-steps model would be undermined. In the next section, we explore alternative ways of interpreting the apparently singular nature of the primary hard-step candidates.

**Box 1 – The Fermi Paradox, the Great Silence, the Drake Equation, Rare Earth, and the Great Filter**

The so-called "Fermi Paradox" is named after physicist Enrico Fermi based on a lunchtime conversation in 1950 in which he asked, "where is everybody?" in the context of the recent reports of UFOs that had been linked to alien spacecraft. The essence of Fermi's question – later formally developed by Hart (*51*) and Tipler (*52*) into what would be called the Fermi Paradox – is that the time to cross the Milky Way galaxy, even in slow ships (taking of order 100 million years, or Myr), is much shorter than the age of the galaxy (of order 10,000 Myr). Any spacefaring species has thus had plenty of time to colonize the Earth – so why do we see no trace of them here on Earth?

The Fermi Paradox is often conflated with the "Great Silence" (*53*), which describes the ostensibly puzzling lack of success of SETI to date to find any signs of technological life *elsewhere* in the Galaxy. This "silence" is purportedly a puzzle because of optimistic numbers one can calculate for our expectation of SETI signals to detect from the Drake Equation, which is a heuristic order-of-magnitude calculation often used to justify SETI efforts. The Drake Equation (*54*) includes a series of terms for the number of potentially life-hosting planets in the Galaxy, the fraction of those planets that give rise to life, intelligence, and technology, and a term for how long such technological life lasts. (Note that most SETI practitioners do not find the lack of success of SETI puzzling at all, given the limited searching that has been done and the very large search space; see for instance Wright, Kanodia, & Lubar (*20*)).

One "solution" to Fermi's paradox and explanation for the Great Silence is that the fractions in Drake's Equation are very small: that is, planets like Earth are so very rare, and the evolutionary contingencies that lead to animal-like life so very unlikely, that despite the huge number of stars in the Galaxy, Earth is the only planet in the Milky Way with such life. This is called the "Rare Earth" argument, made most forcefully by Ward & Brownlee (*55*).

Starting from the assumption that the Fermi Paradox and Great Silence are puzzles demanding a solution, Hanson (*10*) proposed that there exists a "Great Filter": some very unlikely step or set of steps on the road to the development of Galaxy-spanning spacefaring life that prevents it from arising, despite optimistic estimates one might calculate from the Drake Equation.

Hanson identified nine essential steps to the widespread colonization of the Milky Way. Eight of these are in Earth's past, including several of the proposed hard steps mentioned in our review here (Table 1), and an additional first step capturing many other "Rare Earth" terms. Hanson's ninth and final step is an exponential "colonization explosion" that leads to nearly every stellar system becoming inhabited.

The popular appeal of the Great Filter framework is that it allows one to consider whether the hypothetical Great Filter is "ahead of us" or "behind us." That is, it is possible that none of the first eight steps is unlikely, and that human-like technology exists throughout the universe but never spreads because there is something in the galaxy that prevents interstellar colonies from taking hold. This could be a reliable form of destruction of *all* technological species before they become interstellar (such as nuclear war or some other cataclysm). Science fiction also offers the suggestion that it could be an incumbency effect, where another, prior species checks or exterminates any species that attempts to spread beyond its home planet, like a gardener doing weeding or a sort of "galactic immune system."

By the logic of the Great Filter, then, the discovery of ancient life on Mars would eliminate many of the early Great Filter candidates from the list, increasing the chances that the Filter lies in our future and humanity's days are numbered. Alternatively, finding that one or more other Solar System bodies have always been sterile would imply the filter might be behind us, and that we will be the first species to colonize the galaxy. As we demonstrate in the main body of the paper, however, substantive arguments exist against the hard-steps model on which the Great Filter depends.

**Box 2 – How unique are humans?**

The hard-steps model originally concerned the likelihood of the evolutionary appearance of what might be called "human-like" intelligence (i.e., the kind of intelligence required to build a complex, technologically advanced civilization). Such intelligence is often characterized by the following features: tool use and fabrication; problem solving; gaze following; metacognition; a theory of mind; consciousness; prosociality, and language (*56*). In general, the evolution of intelligence is expressed as an increase in cognitive complexity (*57*). A key assumption of the hard-steps project was that human-like intelligence is both special and unique in evolutionary history, and demands a unique explanation. Recent developments in a variety of fields ranging from anthropology to neurobiology raise questions about just how singular human intelligence may be compared to other species.

To argue against the uniqueness of humans begins with the cognitive foundations of intelligence. Even microbial communities have been found to demonstrate rudimentary cognitive capacities through signal transduction and quorum sensing where microbes can sense and communicate environmental changes and coordinate responses (*58*). The characteristic "membrane excitability" that forms the basis of such microbial sensing is the root structure that would later become the basis for neural cells and their capacity to channel and regulate information flow. Further down the evolutionary lineage such nerve cells, even before true brains form, allow metazoans with nerve nets to display simple learning behaviors (*57*). As one reaches more complex metazoans, research now shows that key features of intelligence are distributed across the animal tree of life.

Tool use, for example, is now known to occur in a variety of species including non-primate mammals and non-mammalian animals like crows. The use of puzzle box experiments has shown that problem-solving exists among both primate and non-primate species. The understanding of visual perspective, i.e. gaze following is found in wolves, dogs, monkeys and apes. Thus, as Roth & Docket put it in their comparative study of intelligence among different species "in contrast to a widespread belief even among biologists and anthropologists, we have found no higher cognitive abilities that are unique to humans in a strict sense, i.e., without any precursors" (*56*). The octopus, for instance, uses tools and displays high degrees of intelligence, but is a mollusk whose common ancestor with mammals was presumably a tiny marine worm of some sort, and quite unintelligent. Thus, Earth life shows a continuum of cognitive complexity with the basic tool kit of intelligence established before humans appeared, and most differences shown between species expressible as a matter of degree to which tools in the toolkit have been deployed (*57*).

Thus, while it is clear that humans possess key aspects of intelligence such as innovation in tool use to higher degrees than other species, the basis for seeing "human intelligence" as a "hard step" is uncertain. For instance, it is possible that it evolved multiple times among the hominins or even among primates generally, and that humans "pulled up the ladder" by dominating the niche, driving the other lineages to extinction. Indeed, it is debated whether anatomically modern humans drove other hominin species, notably Neanderthals, to extinction (*59*, *60*), thereby preventing the co-existence of at least two (albeit closely related) animal species with “human-like intelligence” on Earth.

The nature of the singularity of human intelligence is also very unclear. As discussed above, anthropologists struggle to find any single trait that can explain modern humans’ superlative capacity for technology that is unique to humans. Nearly all individual aspects of modern humans that seem necessary to our clearly special technological capabilities – tool use, creativity, abstraction, sense of self, social behaviors, transmission of learned behavior across generations, communication – clearly exist in other lineages. Furthermore, while the question of whether an industrial-type, non-human civilization could have existed much earlier in the Phanerozoic (0.54-0.0 Ga, or billions of years ago) has received little-to-no serious attention from paleontologists, it is arguable that the geochemical, fossil, and material evidence from such a civilization would be either absent or difficult to discern from non-industrial sources in the modern (*61*). Indeed, what will remain of our own civilization in the deep future? Again, it is unclear if the apparent uniqueness of *H. sapiens* should be taken for granted, even though we ostensibly find ourselves “alone” on the Earth today.

**Box 3 – The Major Transitions in Evolution**

The "Major Transitions in Evolution" (MTE) framework was first articulated in two 1995 publications – a book (*62*) and a paper (*63*) – by the evolutionary biologists John Maynard Smith and Eörs Szathmáry. MTE were primarily defined by Maynard Smith and Szathmáry as evolutionary shifts in how "information is stored and transmitted," and served as an attempt to operationalize how "complexity" has accumulated in certain evolutionary lineages over geologic time (*62*, *63*). An additional feature of MTE was that "entities that were capable of independent replication before the transition can only replicate as parts of a larger unit after it" (*63*). For instance, the eukaryotic cell most likely emerged from the union of at least a bacterial symbiont and an archaeal host, both of which were ancestrally capable of living and reproducing on their own (*64*, *65*).

Since 1995, the MTE framework has been updated (*66*) and criticized (*67*), primarily on the grounds that Maynard Smith and Szathmáry's original list of eight major transitions lacked theoretical unity (*28*, *50*). For example, number eight on the original list (*62*, *63*), "primate societies to human societies (language)," fails to meet the criterion of previously free-living entities becoming integrated into higher-level individuals (*28*, *68–70*). To achieve theoretical unity, the focus of the MTE framework has since shifted to the hierarchical (or nested) nature of biological organization (e.g., plants composed of cells, cells composed of organelles, etc.) (*28*, *50*, *71*), with the understanding that a "major transition" constitutes "the emergence of a new population of evolutionary individuals" (*50*) (e.g., eukaryotes from bacteria and archaea). Applying these criteria (Table 2), many MTE – such as the evolution of eusociality in naked mole-rats, or the evolution of coloniality in the Portuguese man o' war – are irrelevant to human origins (*50*). The MTE framework is explicitly not a list of evolutionary milestones required for human intelligence (*50*), or a compilation of evolutionary innovations perceived to be particularly "important" (*50*). As such, it is a category error to substitute the MTE for a list of candidate hard steps (*9*, *12*, *14*), even if certain hard step candidates qualify as MTE (Table 1).

## RESULTS

### Ways around improbability

Evolutionary "singularities" – innovations without parallels in the history of life on Earth – are generally interpreted by evolutionary biologists to reflect the inherent contingency and unpredictability of the evolutionary process, as well as the unlikelihood of evolving the singularity in question (*32*, *72*). In phylogenetic terms, such singularities necessarily belong to "clades" or "monophyletic groups" of organisms (*32*) (Fig. 2a). A group of organisms (e.g., a genus) is monophyletic (a clade) when its members are more closely related to one another than to organisms outside the group (*73*). This pattern results from the group being descended from a common ancestor (*74*) that possessed the characteristics that distinguishes the clade from others (for example, feathers, warm-bloodedness, and flight were all present in the last common ancestor of all living bird species, which together form the clade Aves). In other words, a clade comprises a founding ancestor and all of its descendants, living and extinct. The five hard-step candidates identified in the previous section correspond (and are constrained) to the origin of the five following clades: Life (with a capital "L" (*75*); the last universal common ancestor, LUCA, and all of its descendants); Cyanobacteria (the ancestors of which evolved oxygenic photosynthesis); Eukarya (the last eukaryote common ancestor, LECA, and all of its descendants); Metazoa (the clade containing all animals); and *Homo sapiens*. The fact that our candidate hard steps each correspond to a single clade across the tree of life is what makes them apparently singular events – that is, if they had occurred more than once, then each step would have produced at least two clades (for example, two clades with a eukaryotic grade of cellular organization, or two clades with animal-like multicellularity) (Fig. 2b). However, there are other proposed ways of explaining these phylogenetic patterns.

In 2006, the paleontologist and evolutionary biologist Geerat Vermeij argued – similarly to Barrow & Tipler (*7*) – that the apparent uniqueness of an evolutionary innovation could arise artifactually via information loss (e.g., extinction) over geologic time (*72*). To test this idea, Vermeij first compiled and compared the origination times of purported singularities and innovations that evolved repeatedly in different lineages and found that the examined singularities were significantly more likely to be constrained to the Precambrian (>0.54 billion years ago, or Ga) than the repeated innovations (*72*). Indeed, four of our five candidate hard steps occurred in the Precambrian (Fig. 1). Next, Vermeij compared the sizes of clades that independently, and relatively recently, evolved a repeated innovation (e.g., the labral tooth, a structure used in predation by marine snails), and found that 50-75% of these clades with a fossil record are represented by only five or fewer species (*72*). Together, based on these results, Vermeij concluded that information loss over geologic time could explain the apparent uniqueness of ancient evolutionary innovations when 1) small clades that independently evolved the innovation in question go extinct, leaving no living descendants, and 2) an ancient innovation evolved independently in two closely related lineages, or within a short period of time, and the genetic differences between these two lineages become "saturated" to the point where the lineages become genetically indistinguishable (*72*).

As an illustration of Vermeij's first mechanism, the endosymbiotic origin of plastids (the photosynthetic organelles of eukaryotes, such as the chloroplasts of algae and plants) directly from free-living cyanobacteria has been widely treated as an evolutionary singularity (*76*), and must have happened over 1.03 Ga (*77*). However, this conclusion became complicated with the recognition that the rhizarian amoeba *Paulinella chromatophora* possesses photosynthetic structures called "chromatophores" acquired from a distinct cyanobacterial lineage only 90-140 million years ago (Ma) (*78*, *79*). While eukaryotes with bacterial and archaeal endosymbionts are well known (*80*), some authors have concluded that the chromatophores of *P. chromatophora* qualify as nascent plastids on the basis of protein import from nucleus-encoded genes of chromatophore origin to the chromatophores themselves (*81*, *82*), a hallmark of organelles (*83*). As of now, there are only three known species of photosynthetic *Paulinella* (*84*), fitting the pattern recognized by Vermeij. In contrast, the Archaeplastida, the eukaryote lineage that acquired the more ancient primary plastid, contains over 19,000 microbial species (*85*), and over 500,000 species of land plants (mammals, for comparison, total only around 7,000 species) (*85*, *86*). If these photosynthetic *Paulinella* species had evolved, entered the fossil record, and went extinct in the Precambrian, would modern paleontologists be able to recognize their fossils as representing an independent origin of primary plastids? Almost certainly not. And if these photosynthetic *Paulinella* species had never entered the fossil record in this scenario, then we would have no evidence for them at all. Applying this logic to the earliest fossil record of eukaryotes, many of the difficult-to-assign "eukaryotic" fossils from the middle of the Proterozoic Eon (2.5-0.54 Ga) (*87*) (Figure 1) may not even represent the Eukarya lineage (i.e., total-group eukaryotes) at all, but instead independent fusions of different bacterial and archaeal lineages yielding similar eukaryote-like features, thereby implying multiple examples, or near-examples, of 'eukaryogenesis' (understanding that these organisms would not share an affinity with Eukarya, but instead an organizational "grade") (*72*, *88*). Overall, information loss operating on geological timescales has the power to obfuscate, even erase, evidence for multiple independent origins of ancient innovations, making their apparent uniqueness in the modern day an artifact (*72*).

In addition to information loss, Vermeij also proposed that once an evolutionary innovation becomes established, competition for limited resources may prevent, or eliminate, subsequent origins of similar innovations, thereby favoring a sole surviving lineage (a singularity) without resorting to the improbability of the innovation itself (*72*). Various authors attempting to interpret singularities have argued similarly, each using their own terminology. For example, de Duve referred to this pattern as a "selective bottleneck," and defined it as "any situation where different options are subject to an externally imposed selection process that allows only a single one to subsist" (*32*). Likewise, the paleontologist Simon Conway Morris used the term "incumbency," referring to when organisms with a given innovation "occupy the 'high ground' and are highly tenacious of their niche" (*89*). This general pattern of ecological inhibition has also been called "pre-emptive competition" (*90*), "home-field advantage" (*91*), "prior-residency advantage" (*92*), and "niche preemption" (*93*) and is a kind of "priority effect" (*94*, *95*), where species interactions depend on the order in which species join a community (*93*). For our purposes, such inhibitory priority effects could have actively prevented certain hard-step candidates from evolving more than once – not because these steps were inherently unlikely, but because repeated occurrences were actively inhibited by their first occurrences.

As an illustration, bacteria evolved phototrophy – the metabolic conversion of light energy into chemical energy for growth (*96*) – at least twice: once in the ancestors of the retinal-based phototrophs (the retinalophototrophs), and once in the ancestors of the chlorophyll-based phototrophs (the chlorophototrophs) (*97*). Using a combination of mathematical analyses and modeling exercises, Burnetti & Ratcliff concluded that retinalophototrophs and chlorophototrophs partition phototrophic niche space by optimizing opposite, yet complementary, sides of intrinsic biophysical trade-offs, namely those between efficiency per unit incident light vs. efficiency per unit protein (*98*). Due to this polarizing effect across these trade-offs, neither group of phototrophs is able to occupy the entirety of phototrophic niche space to the exclusion of the other, thereby permitting the persistence of two clades that evolved phototrophy independently. However, both retinalophototrophs and chlorophototrophs in this scenario actively prevent the repeated origin of phototrophic systems like themselves – hence the dynamic maintenance of this "dual singularity" via inhibitory priority effects. This example suggests that the origin of phototrophy, while nearly singular, is not improbable, but fundamentally constrained by these priority effects. Other potential singularities – such as those on our list of candidate hard steps – may indeed be singularities, but not because they are improbable, but because independent origins are actively inhibited via evolutionary incumbency and priority effects, specifically when the innovation in question quickly occupies the available niche space.

Similar to priority effects, a singularity may be maintained not by the improbability of its origin, but by what Shulze-Makuch & Bains call "pulling up the ladder" (*99*). In this scenario, an evolutionary singularity, through its ecological success and environmental impact, ultimately destroys the conditions necessary for its own origin, but not for its persistence (it "pulls up the ladder" after itself). For example, Life – through the origin of oxygenic photosynthesis and the resulting oxygenation of the atmosphere (more below) – destroyed the reducing atmosphere necessary for its initial emergence, but not for its continued existence (*99*). What Shulze-Makuch & Bains describe is essentially what is called niche construction (*100*, *101*), or ecosystem engineering (*102*), in which organisms actively modify and create the physical and geochemical conditions within which they, and other organisms, live, thereby altering the sources of natural selection in their immediate environment. When organisms inhabit environments constructed by the activities of either their direct genetic ancestors or ecological ancestors, they experience what is called ecological inheritance (*100*). For example, every organism on Earth today has inherited the well-oxygenated biosphere ultimately established by early oxygenic phototrophs, regardless of whether they are direct descendants of these phototrophs. Indeed, all five of our candidate hard steps yielded diverse and globally extensive clades that are enormously impactful on Earth's global biogeochemical cycling, so much so that the world they each evolved into no longer exists by virtue of their own behavior and activity (this is perhaps most immediately obvious with oxygenic photosynthesis and life itself – more below). As a result, the singular status of our hard-step candidates may primarily reflect the disproportionate and (so far) irreversible impact their corresponding clades have exerted, and continue to exert, on the Earth system, rather than the inherent improbability of their origin.

In summary, there are at least three ways of interpreting the probability of an evolutionary singularity: 1) the singularity is indeed improbable, the product of contingency, hence why it failed to evolve repeatedly in disparate lineages (consistent with it being a hard step); 2) the singularity is probable, but remains a singularity via evolutionary priority effects and/or biospheric niche construction; and 3) there is no evolutionary singularity – information loss over geologic time has created the illusion of one, thereby increasing the probability of the innovation in question. Looking at our list of candidate hard steps, it is possible that all five represent a combination of either scenarios 2 or 3, which would, in turn, eliminate them all as hard steps.

The apparent singularity of our hard-steps candidates have already been interpreted through the lens of evolutionary priority effects and information loss. Starting with abiogenesis, Raup & Valentine (*103*) estimated that life could have originated at least 10 separate times with only one clade (LUCA and its descendants) surviving to the modern day – an illustration of information loss. Alternatively, it has also been suggested that life could have originated once and, after becoming globally established, prevented subsequent origins of life by competitively excluding nascent lifeforms into extinction (*103*, *104*) or, ultimately, by oxygenating the atmosphere – an illustration of evolutionary priority effects or "pulling up the ladder."

Similarly, oxygenic photosynthesis may have evolved only once (in the ancestors of Cyanobacteria), but, as Lenton & Watson (*14*) speculated, "once the ancestral cyanobacterium had evolved and had become established, it [may have] suppressed any tendency for other potential oxygen producers to evolve, by out-competing them before they had time to get the biochemistry right."

With respect to the eukaryotic cell, Vermeij proposed that multiple metabolically analogous, yet phylogenetically unique, bacterial-archaeal couplings – like the one that ultimately lead to LECA (*64*, *65*, *105*) – may have existed before LECA originated, with only the descendants of LECA (i.e., crown-eukaryotes) surviving to the modern day (*72*). This scenario (information loss creating the illusion of a single origin of eukaryotes) was similarly suggested by Booth & Doolittle (*88*), although they criticized the idea that crown-eukaryotes actively outcompeted their "near-eukaryotic sister lineages" to extinction as "ecologically naive."

Next, the origin of multicellularity in animals (or metazoans) arguably fails as a singularity, as at least four other extant eukaryote lineages exhibit comparable grades of "complex" multicellularity (*46*). Claims that metazoan multicellularity is an evolutionary singularity primarily concern cellular and ecological traits present in animals and their direct unicellular ancestors, but absent from other complex multicellular groups and their respective ancestors. Notably, animal cells are capable of phagocytosis ("cell swallowing"), which allows them to internalize other cells for nutrition, immune system functioning, and development (*47*, *106*). In contrast, this capacity is absent in the other complex multicellular clades (e.g., land plants, mushroom-forming fungi) due to their possession of cell walls (*106*). Why animal-like (phagocytic) complex multicellularity evolved only once – or only has one surviving example – is unclear. Outside of intrinsic difficulty (*47*, *48*), animals may have quickly saturated available niche space in ways that complex multicellular phototrophs (e.g., land plants, kelp) and fungi did not. Alternatively, other examples of animal-like complex multicellularity may have originated in other non-metazoan lineages, but have since gone extinct (*48*).

Last on our list, the uniqueness of human origins is addressed in Box 2. Together, there are reasonable, yet poorly explored, alternatives to the face-value interpretation of our hard-step candidates as both unique and unlikely events, raising the possibility that these transitions were all more probable relative to the lifespan of the biosphere.

If none of our hard-step candidates were intrinsically improbable relative to the lifespan of Earth's biosphere (that is, there are no hard steps), then why do they have the temporal distributions that they do (Fig. 1)? In other words, if these evolutionary events were all likely to occur in the externally allotted time, then why do we find ourselves so close (1.0 ± 0.5 Gyr) to the outer edge of Earth's habitability? Why do we not observe ourselves on a much younger Earth? Going back to the formulation of the hard-steps model itself, Carter explicitly rejected $t_i \ll \tau_0$ on the grounds that 1) the Earth is too old to reconcile with this scenario, and 2) because "it is hard to think of any particular reason why our arrival should have been greatly delayed relative to the intrinsically expected time" (*1*). However, it is not hard at all for deep-time paleontologists, geochemists, and Earth system modelers to think of reasons why our arrival could have been so "greatly delayed." Indeed, the hard-steps model famously ignores perhaps the prime determinant of macroevolutionary timescales on Earth – the Earth system itself (*14*, *30*).

**Increasing habitat diversity over geologic time**

Dismantling the singular nature of these candidate hard steps (due to information loss), or explaining their singularity outside of improbability (due to priority effects and/or niche construction), opens the possibility that these traits and transitions each evolved in response to the crossing of key global environmental thresholds – the opening of "permissive environments" (*107*). Generally, singular (or monophyletic) transitions are interpreted to reflect the rare and improbable overcoming of intrinsic constraints, such as structural, metabolic, or genetic bottlenecks that prevent multiple lineages from evolving the same innovation (*48*, *108*, *109*). In contrast, the lifting of environmental barriers is generally predicted to yield "polyphyletic radiations," in which multiple preadapted lineages simultaneously evolve a given innovation in response to the collective passing through an environmental bottleneck (*62*, *108–110*). Note, in these environmental threshold scenarios, environmental change does not itself cause or explain the evolutionary innovation in question, but instead represents the removal of an external constraint that had previously prevented the innovation from evolving (*111*). Therefore, "permissive environment" scenarios explain the timing of evolutionary innovations (that is, why these innovations occurred when they did and not earlier) – they are not mechanistic explanations for the innovations themselves (*111*). In light of the arguments presented in the previous section, it could be that some of our hard-step candidates evolved polyphyletically in response to environmental change, but this phylogenetic pattern has since been lost or obscured via billions of years of information loss. At the same time, some of these candidate hard steps may not have evolved polyphyletically in response to global environmental change, but only because the first lineage to realize the innovation (sometime after the environment became permissive) rapidly filled the available niche space to the extent that it actively prevented additional "prime" lineages from repeating the evolutionary innovation via priority effects and/or ecosystem engineering. To test the idea that global environmental change over Earth history controlled when our hard-step candidates evolved, two lines of evidence need to be explored: 1) the environmental requirements of each hard-step candidate; and 2) when these environmental requirements first became met by the global environment.

Far from being the static setting Carter originally envisioned (*1*), Earth's surface environment has radically and irreversibly transformed itself over its 4.6-billion-year history, primarily as a consequence of life itself (Fig. 3) (*14*, *107*, *112*). Perhaps the most commonly invoked environmental variable for explaining why certain evolutionary innovations occurred when they did (and not earlier) is atmospheric oxygen ($O_2$) (*30*, *113–117*), which, in turn, is a consequence of biological evolution, specifically photosynthetic $O_2$ production (*118*, *119*). While the details remain debated, Earth historians generally agree that atmospheric oxygen evolved in the following broad stages (*118*, *120*, *121*): 1) Earth's atmosphere was initially anoxic (that is, $O_2$-free) until no later than 2.4-2.2 Ga (*122*), during the so-called Great Oxidation Event (or GOE) (*123*, *124*) when the partial pressure of atmospheric $O_2$ ($pO_2$) irreversibly exceeded 0.001% ($10^{-5}$) of present atmospheric levels (PAL) (*125*, *126*); 2) following the GOE, $O_2$ remained a stable feature of the atmosphere (>$10^{-5}$ PAL), albeit at non-modern levels – usually constrained to a minimum of ~1-10% PAL (*127–132*), although 0.05-1% PAL is also possible (*133*, *134*); 3) following an interval of dynamic and increasing $pO_2$ levels (*135–137*), $pO_2$ stabilized near present values (~100% PAL, or ~21% of the atmosphere by volume) by 420-400 Ma during the so-called Paleozoic Oxygenation Event (POE) (*135*, *138–140*), and have been maintained near these levels ever since (Fig. 3). By this broad estimation, up to 52% of Earth's history had elapsed before atmospheric $O_2$ stabilized above trace amounts, and up to 91% had elapsed before $pO_2$ stabilized at near-modern levels, emphasizing the non-uniformitarian nature of Earth's redox landscape over the last 4.6 billion years.

With respect to our candidate hard steps, the last three (the origin of eukaryotic cells, animal multicellularity, and humans) on Earth all have absolute environmental $O_2$ requirements, and must have been precluded by the global environment before their minimum $O_2$ requirements were met. Starting with *H. sapiens*, the lower $O_2$ limit for long-term human habitation – based on $pO_2$ at the highest-altitude human settlements (*141–143*) – is estimated to be 53-59% PAL $O_2$. While these minimum $O_2$ levels may have been transiently met in the Neoproterozoic Era (1.0-0.539 Ga) and the earliest Paleozoic Era (539-252 Ma) (*136*, *143*), they apparently did not become reliably established until the POE ca. 420-400 Ma (*138–140*), implying that Earth's atmosphere has only been conducive to long-term human habitation for the last ~9% of its total history – not 100% as Carter assumed (*1*). Furthermore, it is estimated that open-air ignition and maintenance of combustion requires at least 87.8% PAL $O_2$, implying that industrial civilization based on combustion technologies (another commonly proposed hard step, Table 1) requires $pO_2$ levels beyond those required for human life itself (*144*). That said, the minimum $pO_2$ requirements for both humans and combustion-based industry were likely met around the same time by the end of the POE.

Next, the lower $O_2$ limit for the origin of animal multicellularity is controversial and essentially unknown (*49*, *145*, *146*). Geobiologists have theoretically (*147*, *148*) and experimentally (*149*, *150*) constrained the minimum oxygen requirements of animals to around 0.1-1% PAL $O_2$ (*151*), based primarily on the physiology of living animals. By some estimates, $pO_2$ already exceeded this threshold by the middle of the Proterozoic Eon (2.5-0.54 Ga), hundreds of millions of years before animals entered the fossil record (*127*, *130*). By other estimates, however, mid-Proterozoic $pO_2$ was around this threshold, or perhaps even below it, until $pO_2$ rose near the time that animals likely originated (*133*, *152–154*). Therefore, it is debated whether animal multicellularity originated only shortly after its minimum $O_2$ requirements became established (*49*, *117*). Recently, experimental evolution studies on snowflake yeast – a model system of simple multicellularity – investigated the growth of multicellular yeast clusters under three different $pO_2$ levels (0%, 27%, and 72% PAL) (*155*). In these experiments, yeast clusters significantly increased in size under anoxia and 72% PAL $O_2$ (using yeast variants incapable of respiring $O_2$ and variants only capable of respiring $O_2$, respectively). In contrast, yeast clusters in the 27% PAL $O_2$ treatment remained closer to their original size, especially for yeast variants that could only respire $O_2$. While 27% PAL $O_2$ clearly exceeds the minimum $pO_2$ estimates for the mid-Proterozoic (Fig. 3), these results raise the possibility that low, non-zero $O_2$ concentrations in the Proterozoic ocean could have actively suppressed the origin of animal multicellularity, or at least the size of the earliest multicellular animals. However, the applicability of these results to the origin of animal multicellularity is unclear, since the single-celled ancestors of animals – unlike yeast – likely possessed flagella (whip-like structures that generate water currents) (*106*) and were facultative aerobes (capable of metabolizing with and without $O_2$) (*156*). Theory – inspired, in part, by experiments conducted with flagellated multicellular algae (*157*) – suggests that the simple multicellular ancestors of animals could have actively bypassed diffusive $O_2$ transport and enhanced internal $O_2$ delivery via the coordinated beating of their surface flagella (*110*), perhaps permitting colony sizes unobtainable by diffusion-limited organisms (like snowflake yeast) under low $O_2$. Overall, it remains debated whether low $pO_2$ levels in the mid-Proterozoic directly prevented the origin of multicellular animals (*49*, *117*), primarily due to poor constraints on the both ancient $pO_2$ levels and the $O_2$ levels needed to permit the evolutionary origin of animal multicellularity. Nevertheless, there is universal agreement that environmental $O_2$ was a necessary precondition for animal life on Earth (*49*, *108*), meaning that the origin of animal multicellularity was likely prohibited by the global environment before the GOE (the first 52% of Earth history).

Not only did the origin of animal multicellularity require sufficient $O_2$ levels, it required the origin of the modern eukaryotic cell (LECA). When LECA emerged in the Proterozoic Eon remains unclear (*87*) (Fig. 1), but the $O_2$ requirements for eukaryogenesis appear more straightforward. Although many living eukaryotes have lost their ancestral capacity to respire $O_2$ (*156*), eukaryogenesis is thought to have required 0.001–0.4% PAL $O_2$ (*158*), based on the $O_2$ requirements for aerobic respiration (*159–161*) and steroid synthesis (*162*) – two $O_2$-dependent processes most likely present in LECA. According to these constraints, the minimum $O_2$ requirements for eukaryogenesis have been met since the GOE, and perhaps even earlier depending on when oxygenic photosynthesis emerged (*158*). Overall, with respect to our three $O_2$-requiring hard-step candidates, the origin of the eukaryotic cell and the origin of animal multicellularity were most likely excluded by the global environment – with respect to $pO_2$ alone – until ~2.2 Ga (the first 52% of Earth history), while the origin of *H. sapiens* and industrial society were excluded until ~0.40 Ga (the first 91% of Earth history).

The timing of our first two hard-step candidates – abiogenesis and the origin of oxygenic photosynthesis – cannot be explained in terms of changing $pO_2$ levels. For one, the origin of life necessarily occurred under (and likely required) anoxia (*163*, *164*), as it predated the origin of oxygenic photosynthesis, the only considerable source of $O_2$ to the atmosphere (*118*, *119*), which, in turn, must have occurred under anoxia. However, other global environmental factors have been invoked to explain the timing of these evolutionary events. First, the earliest point at which the Earth became habitable has been called the "habitability boundary," and is currently constrained to between ~4.5 and ~3.9 Ga, based on temporal estimates for when liquid water oceans formed on one end and potentially sterilizing meteorite impacts ceased on the other (*165*). Comparing this range to the earliest purported isotopic evidence of metabolism in sedimentary rocks ~3.7 Ga (*166*) suggests that life could have arisen within a permissive window as brief as 200 Myr or as long as 800 Myr (*165*). In either scenario, the origin of life – like our other hard-step candidates – likely awaited the establishment of global environmental conditions that were not immediately present upon Earth's formation.

Next, there is a long-standing debate concerning sea surface temperatures in the Archean Eon (4.0-2.5 Ga). The oxygen (*167*) and silicon (*168*) isotope composition of cherts, as well as the thermostability of resurrected proteins (*169*, *170*), have all been interpreted as reflecting Archean sea surface temperatures >70°C. Meanwhile, the upper temperature threshold for modern cyanobacterial growth is reliably constrained to 70-73°C (*171*), and cyanobacteria are predicted to have originated at temperatures approximating 64°C according to resurrected elongation factor proteins (*169*). Together, these lines of evidence have led certain Earth historians to predict that the cooling of the Archean climate controlled when cyanobacteria both originated (*172*) and flourished (*173*). However, if the Archean climate was more mild (0-40°C), as many climate modelers predict (*174*, *175*), then global environmental factors other than sea surface temperature – such as the growing extent of global subaerial landmass and freshwater availability (*176*) – may have dictated when oxygenic photosynthesis originated.

Like the proposed "habitability boundary" for life in general (*165*), any conceivable hard step must have its own respective habitability "window," within which the global environmental conditions necessary for both the origin and persistence of these innovations are met and sustained. It appears that unidirectional changes in Earth's surface environment (Figure 3) – notably, but not exclusively, the protracted oxygenation of the atmosphere (*121*) – have increased the diversity of habitats over Earth history (*172*), permitting wider varieties of organisms as the Earth system evolves. The modern Earth, with 100% PAL $O_2$ and $O_2$-rich deep oceans, permits organisms (e.g., *H. sapiens*, blue whales, etc.) and ecosystems (e.g., coral reefs, rainforests, etc.) that were simply impossible for the vast majority of Earth's history. The modern Earth, however, also preserves anaerobic ($O_2$-independent) life, as $O_2$-free environments (e.g., marine sediments, marine water columns underlying productive surface waters) have persisted uninterrupted from life's origin to the modern day, meaning that habitat diversity with respect to $O_2$ tolerance has increased over geologic time (rather than aerobic life simply replacing anaerobic life), and that functional diversity has accumulated over Earth history. Furthermore, while humans descend from the earliest metazoans and eukaryotes, and require the $O_2$-rich atmosphere ultimately created by our ecological ancestors (previous oxygenic phototrophs), humans also depend on the continued existence of these groups for long-term habitation (e.g., nutrition, the maintenance of modern $pO_2$ levels, etc.), meaning that the window of human habitability necessarily overlaps with the respective windows of the remaining hard steps.

While we focus on $pO_2$ here for the sake of brevity, other changing surface variables implicated in driving evolutionary timescales, or permitting the origin and success of particular hard-step candidates, have been proposed (and debated), and it is likely that the confluence of numerous surface variables and events explains macroevolutionary timescales better than any single variable in isolation. These proposed variables and events include: increasing primary productivity (*177*) (Fig. 3), increasing atmospheric ozone ($O_3$) levels (*178*, *179*), increasing nutrient availability (*180*, *181*), decreasing sea surface temperature (*172*, *182*), decreasing ocean salinity (*183*), decreasing $pCO_2$ (*184*), Snowball Earth glaciations (*185*, *186*), landscape dynamics (*187*), and the evolution of modern-style plate tectonics (*188*). Considering all of these factors together, Earth's surface environment and life may have co-evolved in such a way that our candidate hard steps evolved and radiated when they did (and not much earlier) due to the sequential and cumulative lifting of different global environmental constraints over geologic time, which together promoted increasing habitat diversity over Earth history. This possibility was absent from Carter's original formulations (*1*) – although it was later raised in subsequent criticisms (*14*, *22*, *189*) – resulting in Carter's rejection of $t_i \ll \tau_0$ and preference for $t_i \gg \tau_0$, which prompted the proposal of hard steps in the first place. In contrast to Carter's logic, the co-evolution of Earth's surface environment and life may explain the temporal distributions of our candidate hard steps (Figure 1) – that is, why our arrival was "greatly delayed relative to the intrinsically expected time" (*1*) – explaining why we find ourselves close to upper bound of Earth's habitability window without invoking the existence of hard steps.

**Planetary constraints and environmental trajectories**

The above scenario, in which unidirectional changes in Earth's surface environment drive ever-increasing levels of habitat diversity, is predicated upon a planetary body inherently capable of physically and chemically accommodating such trends (both in general and in the available time). As such, it is possible that fixed physical and chemical conditions – such as Earth's initial mass, composition, mantle redox state, and orbital distance – could serve as candidate "try-once" steps (that is, non-biological "steps" required for our existence that immediately succeeded or failed with no opportunity to recover from failure) (*190*). On habitable worlds without the same starting conditions as Earth, global transitions like the oxygenation of the atmosphere (analogous to Earth's GOE) may occur more slowly (relative to Earth) or not even at all (*21*) – even following a relatively quick and probable origin of oxygenic photosynthesis. For example, $O_2$ production might never exceed $O_2$ consumption on these worlds, preventing the establishment of Proterozoic- and Phanerozoic-type biospheres and the corresponding evolution of human analogues. Or perhaps atmospheric oxygenation does eventually occur on a subset of these inhabited worlds, but too soon before the irreversible extinction of their respective biospheres to permit human-like life. At the same time, however, planetary transitions like atmospheric oxygenation could conceivably occur even more rapidly on other inhabited worlds than they did on Earth provided the requisite starting conditions (*21*). In other words, the pacing of global-environmental change (Fig. 3) and biological evolution (Fig. 1) on Earth may have actually proceeded much more slowly compared to rates potentially achieved elsewhere, thereby explaining the coincidence between $t_e$ and $\tau_0$ without invoking hard steps (*21*, *23*). Overall, while planetary starting conditions inescapably constrain biospheric evolution, it nevertheless remains unclear how typical Earth's starting conditions are among inhabited worlds, and whether the environmental-biological evolution of Earth has proceeded at a rapid, typical, or slow pace (*21*).

In addition to a world intrinsically capable of accommodating the surface trends needed to ultimately permit human-like life (Figure 3), our proposed alternative to the hard-steps model also requires that these trends and transitions are themselves predictable or probable. Whether this situation applies to the Earth is unknown, but a case for the inherent likelihood of environmental transitions (within the externally allotted time) can arguably be made once again using the GOE as an example. While it remains debated whether the origin of oxygenic photosynthesis predated the GOE immediately (geologically speaking) (*191*) or by hundreds of millions of years (*121*), both end-member possibilities suggest that the GOE was an unavoidable consequence of oxygenic photosynthesis. In the former case, photosynthetic $O_2$ production overwhelmed $O_2$ sinks like atmospheric $CH_4$ and $H_2$ on geologically rapid timescales (~$10^5$ years) (*191*). In the latter case, this tipping point awaited factors that are still unclear (*121*), but may have involved Earth's tectonic evolution – for example, increased subaerial volcanism (*192*), or changes in the composition of continental crust (*193*) – which, in turn, is ultimately driven by the secular cooling of Earth's mantle (*194*). Therefore, following the origin of oxygenic photosynthesis, both scenarios suggest that the GOE required only sufficient time (although the required time differs by three orders of magnitude). Indeed, even in the absence of specific trigger events like those dictated by tectonic evolution, Earth's broader stepwise oxygenation (Fig. 3c) has been modeled as the result of dynamics inherent to Earth's biogeochemical cycling, meaning that once oxygenic photosynthesis evolved, the unidirectional oxygenation of Earth's atmosphere – including the eventual rise to modern levels – was indeed only a matter of time (*140*, *195*). Together, if oxygenic photosynthesis was not a hard step, as explored above, then the trajectory of atmospheric $O_2$ over Earth's history (Fig. 3c) may have been similarly predictable and probable, rather than a series of intrinsically unlikely global-environmental transitions (or "geologic" hard steps). This same intrinsic likelihood may also apply to other surface conditions that have varied over geologic time, such as primary productivity (Fig. 3b) (*177*), although a systematic review of each surface variable and its controls over Earth history is beyond the scope of this paper. Alternatively, if global-environmental transitions like the GOE and the rise to modern $pO_2$ levels on Earth were inherently unlikely in the available time, then a version of the hard-steps model emphasizing geologic hard steps would persist, even in the absence of biological hard steps, as proposed here.

**The ongoing search for exoplanetary biosignatures and technosignatures**

Ongoing and near-future exoplanet observatories will focus on characterizing the atmospheric composition of planets beyond our Solar System (*196*). If the origin of life is indeed a hard step, then atmospheric (or gaseous) biosignatures – volatiles either produced directly or indirectly by life and/or biogeochemical cycling (*196*) – are predicted to be rare among Earth-like exoplanets (geologically active, rocky planets with $N_2$-$CO_2$-$H_2O$ atmospheres and surface temperatures supportive of liquid water). If the origin of life is not a hard step, but the origin of oxygenic photosynthesis is (or, from a geologic perspective, atmospheric oxygenation is), then the detection of Archean-like biospheres might be common among Earth-like planets, while Proterozoic- and Phanerozoic-type atmospheric spectra will be rare or unobserved. Alternatively, if our proposed model is correct, then Archean-, Proterozoic-, and Phanerozoic-type atmospheric spectra will be common or even universal among Earth-like exoplanets, with exoplanetary age perhaps determining which type is observed (going from youngest to oldest, respectively). Furthermore, if there are indeed no hard steps (biological or geologic) leading to the evolution of modern human-like civilization, then planetary technosignatures – remotely observable signatures of extraterrestrial technology (*197*) – are predicted to be common on Earth-like planets of sufficient age. Therefore, we propose that such exoplanetary characterization in the near future will be able to distinguish between the theoretical possibilities outlined here.

**DISCUSSION – An alternative model of biospheric evolution**

The hard-steps model was originally motivated to explain the temporal coincidence between the age of the Earth as it now appears to us ($t_e$, the timing of our emergence, 4.6 Gyr after Earth's formation) and the upper limit of Earth's habitability window ($\tau_0$, now estimated to ~5.6 Gyr after Earth's formation). In other words, why does the timing of human origins so closely approximate the extinction of the biosphere? Carter's preferred solution was that the average intrinsic time for the evolution of "intelligent observers" ($t_i$) well exceeds the lifespan of the biosphere ($\tau_0$), inspiring the notion of hard steps to explain why humans (or human analogues) are so unlikely to evolve within this timeframe (*1*). Alternatively, we raise the possibility that there are no hard steps (despite the appearance of major evolutionary singularities in the universal tree of life) (*72*), and that the broad pace of evolution on Earth is set by global-environmental processes operating on geologic timescales (i.e., billions of years) (*30*). Put differently, humans originated so "late" in Earth's history because the "window of human habitability" has only opened relatively recently in Earth history (Fig. 4). This same logic applies to every other hard-steps candidate (e.g., the origin of animals, eukaryogenesis, etc.) whose respective "windows of habitability" necessarily opened prior to humans, yet sometime after the formation of the Earth. In this light, biospheric evolution may unfold more deterministically than generally thought, with evolutionary innovations necessarily constrained to particular intervals of globally favorable conditions that opened at predictable points in the past, and will close again at predictable points in the future (Fig. 4) (*198*). Importantly, Carter's anthropic reasoning still holds in this framework: just as we do not find ourselves living before the formation of the first rocky planets, we similarly do not find ourselves living under the anoxic atmosphere of the Archean Earth (Fig. 4).

To test the framework proposed here, at least two major areas of research need to be advanced. First, the singular (or unique) status of evolutionary innovations required for human existence, such as our hard-step candidates (Fig. 1), needs to be more explicitly questioned (*72*, *98*). That is, are these innovations truly singular in Earth history, and if so, are they singular because they were intrinsically difficult and improbable, or because evolutionary priority effects and/or ecosystem engineering have prevented repeated origins? Next, the environmental conditions (e.g., $O_2$ availability, ambient temperature, pH, salinity, primary productivity, etc.) required for each candidate hard step need to be more rigorously and exhaustively defined. These requirements then need to be compared to paleoenvironmental reconstructions and Earth system models to determine when these conditions first became established in Earth's past and when they will likely disappear in Earth's future. The target of this research is a more exhaustive version of Fig. 4 encompassing a greater number of Earth system parameters (not just $pO_2$) applicable to each candidate hard step (not just humans).

The implications of our proposed alternative to the hard-steps model extend well beyond assessing the likelihood of "human intelligence" on Earth. This framework can be applied to any evolutionary innovation in Earth's past, whether or not the innovation in question led to the origin of *H. sapiens*. Indeed, this framework raises the possibility that biospheric evolution generally proceeds in a coarsely deterministic or predictable fashion, governed by long-term biospheric trends like increasing habitat diversity in response to unidirectional changes in Earth's surface environment (Fig. 3) (*172*, *198*). Not only would these trends and processes apply to the Earth through time, but their analogues may apply to other inhabited Earth-like worlds in the Universe. In this sense, not only would the evolutionary origin of *H. sapiens* be more inherently probable than Carter predicted (*1*), but so would the evolutionary origins of human analogues beyond Earth.

**Acknowledgements**
DBM acknowledges the influence of David Schwartzman. This manuscript benefited from the helpful comments of Richard Boyle, Anthony Burnetti, Niklas Döbler, Martin Dohrmann, and Warren Francis, as well as from discussions with Stuart Daines and Timothy Lenton. JTW thanks James Endres Howell for helpful discussions. Funding: DBM was supported by the Deutsche Forschungsgemeinschaft (DFG) through Project OR 417/7-1 (granted to William D. Orsi). We gratefully acknowledge support from the NASA Exobiology program under grant 80NSSC20K0622. The Center for Exoplanets and Habitable Worlds and Penn State Extraterrestrial Intelligence Center are supported by Penn State and its Eberly College of Science. Author contributions: DBM, JLM, AF, and JTW conceived the study, and DBM wrote the paper with contributions from JLM, AF, and JTW. All authors declare that they have no competing interests.

## Figure Captions, Tables, and Table Captions

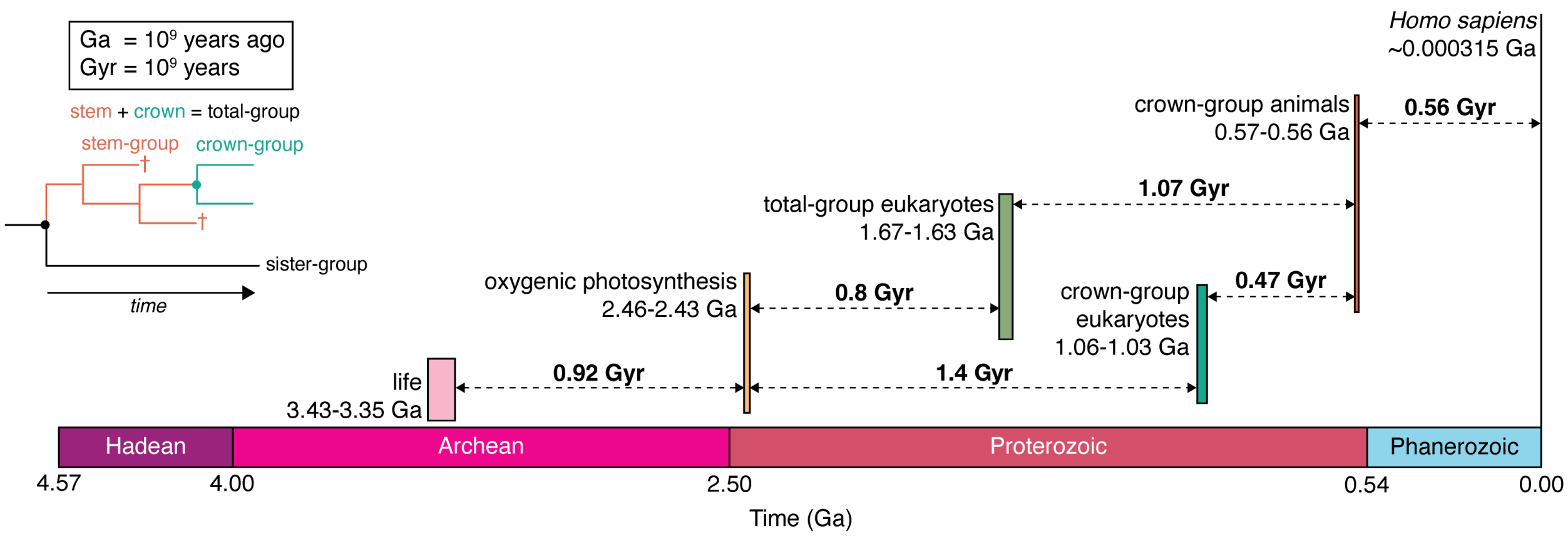

**Figure 1. The temporal distribution of our candidate hard steps.** The vertical colored bars represent the earliest unequivocal evidence for each candidate hard step in the geologic record with widths spanning the upper and lower age constraints (bar lengths are arbitrary). While there exist more contentious geochemical and molecular clock estimates for these steps that would place them each farther back in time, we have chosen the least controversial evidence to produce the most conservative timeline possible. Therefore, each candidate hard step necessarily preceded, but occurred no later than, their displayed dates, and the incorporation of other lines of evidence would necessarily shift the origin of each step back in time to varying degrees. The time intervals separating adjacent steps were calculated using the minimum age constraints only and are displayed in bold and expressed in billions of years (Gyr). With respect to the eukaryotic fossil record, there is ongoing uncertainty concerning when the last eukaryote common ancestor (LECA) evolved (*86*), which marks the completion of "eukaryogenesis" (*196*). Specifically, it remains unclear whether LECA emerged hundreds of millions of years before the oldest eukaryotic-grade fossils (1.63-1.67 billion years ago, or Ga), or hundreds of millions of years after (to use the two end-member scenarios) (*86*). In order to explore both scenarios, we display: 1) the oldest fossil evidence for recognizable crown-eukaryotes (1.06-1.03 Ga), which designates all eukaryotes, extant and extinct, descended from LECA, and 2) the oldest fossil evidence for total-eukaryotes (1.67-1.63 Ga), which comprises both crown-eukaryotes and now-extinct eukaryote lineages that diverged prior to LECA (stem-eukaryotes). A cladogram depicting the concepts of total, stem, and crown groups is displayed on the left, with † designating extinct stem-lineages. Data sources: oldest evidence for life (*197–200*); oxygenic photosynthesis (*201*, *202*); total-eukaryotes (*203*); crown-eukaryotes (*76*); crown-metazoa (*204*); and *Homo sapiens* (*205*).

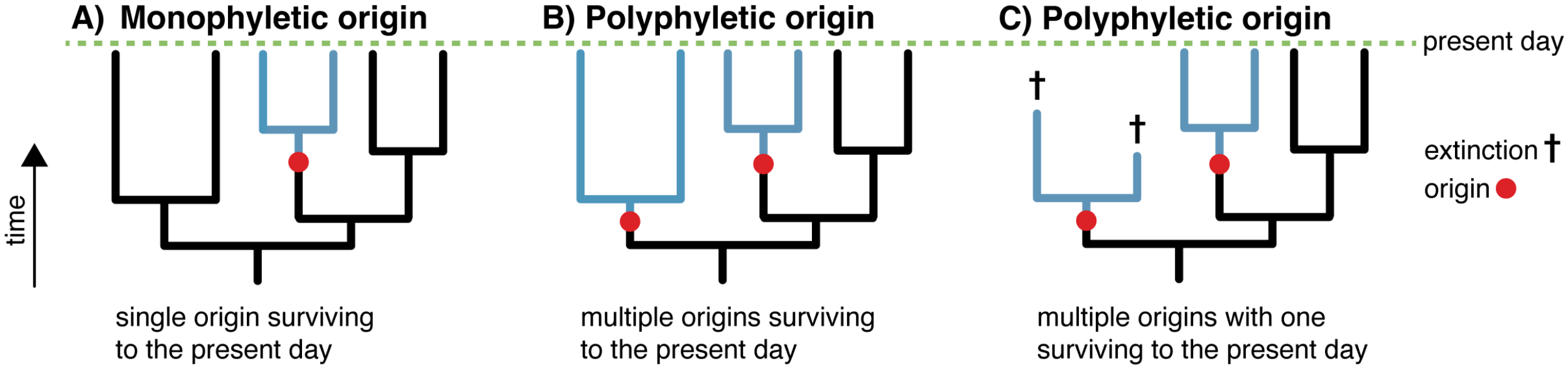

**Figure 2. Phylogenetic comparisons between a single origin vs. multiple origins of an evolutionary innovation.** A) The phylogenetic pattern reconstructed when a given evolutionary innovation is constrained to a single living clade (monophyletic group), the result of a single origin (designated by the red dot); B) The phylogenetic pattern reconstructed when a given evolutionary innovation is found in two different living clades, the result of two independent origins; C) The phylogenetic pattern reconstructed when an evolutionary innovation is constrained to a single living clade, but as the result of the extinction of lineages that had independently evolved the innovation.

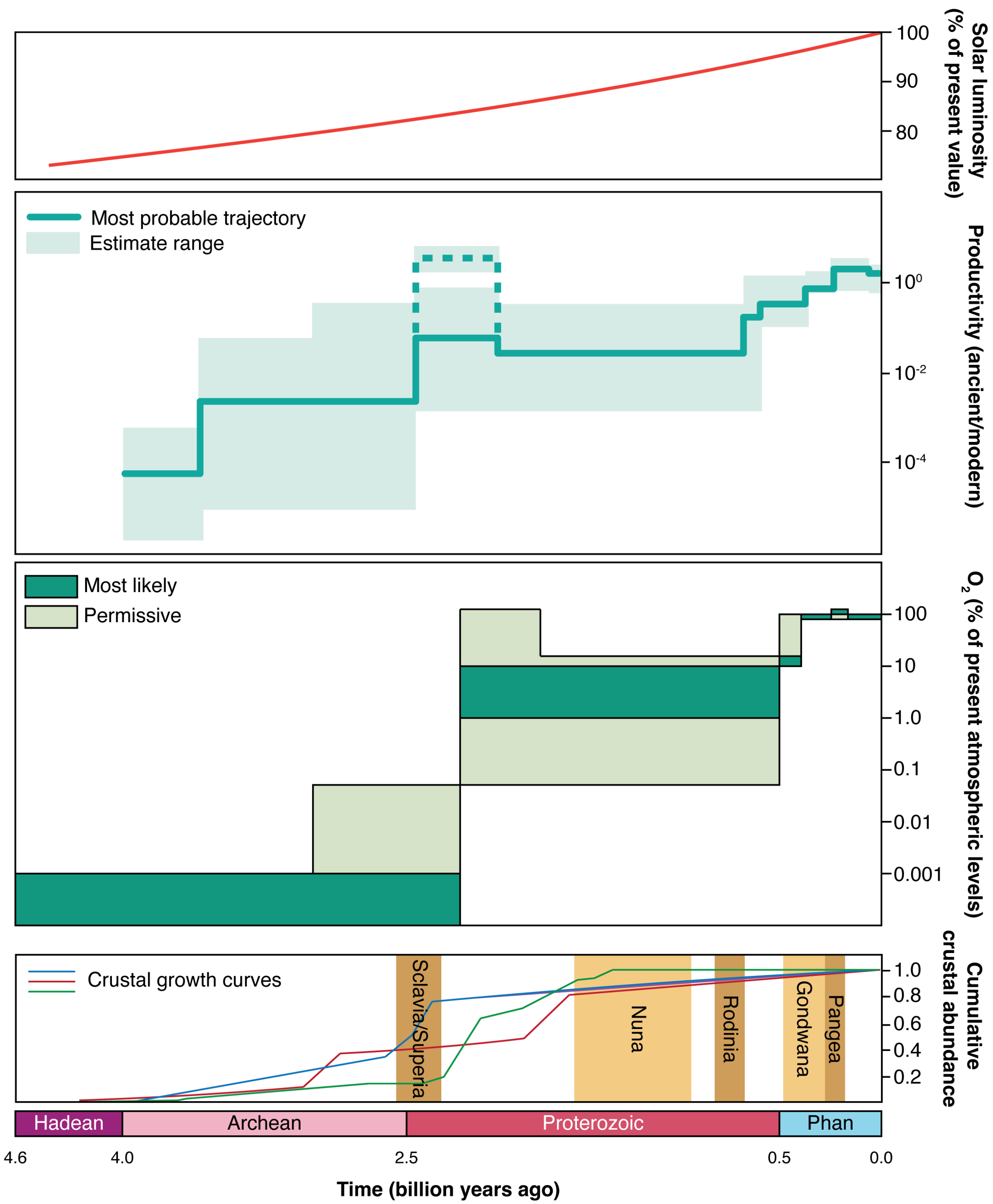

**Figure 3. Unidirectional changes in Earth's surface environment over geologic time.** A) Increasing solar luminosity (expressed as percentage of modern values) since the origin of the Earth (*206*). B) Increasing primary productivity (expressed as the ratio between ancient and modern levels) over Earth history, using values from Figure 1A from Crockford *et al*., 2013

(*176*). C) Increasing atmospheric $O_2$ (expressed as the percentage of present atmospheric levels) over geologic time, taken from Figure 3 of Mills *et al*., 2022 (*157*). D) Evolution of the geosphere (crustal abundance, supercontinental cycles) over geologic time, taken from Figure 1 from Crockford *et al*., 2019 (*207*) and references therein. Overall figure design and content was inspired by Figure 1 from Crockford et al., 2019 (*207*). Phan = Phanerozoic Eon, 539-0 million years ago.

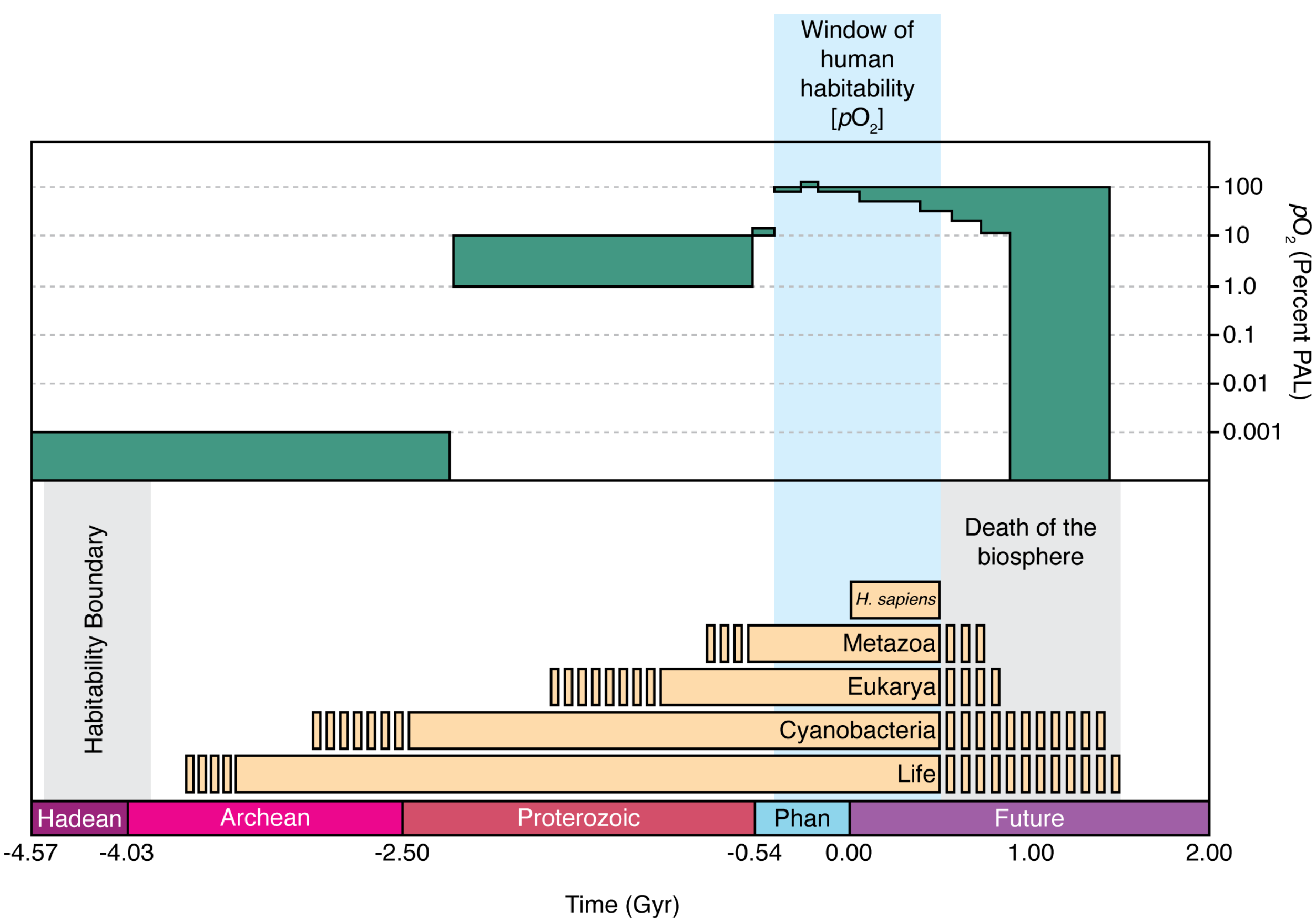

**Figure 4. The total lifespan of the biosphere.** The lifespan of the biosphere is necessarily constrained between the onset of Earth's habitability (the "habitability window," constrained between ~4.5 and ~3.9 billion years ago) (*164*) and its end (the extinction of all life, constrained to ~1.0 ± 0.5 billion years into the future) (*40–42*). The temporal distributions of the five extant clades corresponding to each of our candidate hard steps (Figure 1) are displayed by the horizontal bars, with dashed segments representing uncertainties surrounding the timing of the origin (left) and eventual extinction (right) of each group. The timing of extinction for each group is purely schematic, following the general prediction that declining $pO_2$ in the future (as well as other factors not displayed here, such as rising sea surface temperatures) will drive these groups to go extinct in the reverse order of their appearance (*42*). The "window of human habitability [$pO_2$]," represented by the blue vertical bar, approximates the interval of Earth's total history (past and future) where $pO_2$ exceeds the threshold necessary to support long-term human habitation (53-59% PAL $O_2$) (*140–142*). The atmospheric $O_2$ curve (green) was modified from Ozaki & Reinhard 2021 (*208*). Gyr = billions of years; $pO_2$ = partial pressure of atmospheric $O_2$; PAL = Present Atmospheric Levels

**Table 1. Compiled hard-step candidates.** A non-comprehensive list of the published hard-step candidates. Publications that resort to the major transitions in evolution as proposed by Maynard Smith and Szathmáry (*61*, *62*) – e.g., Hanson 1998 (*9*) and Watson 2008 (*12*) – are excluded here and discussed in Box 3. For comparison, though, the hard steps listed here that correspond to major transitions in evolution (Box 3) as identified by Herron 2021 (*49*) are designated by the † symbol. For simplicity, differently phrased and closely related hard steps were combined into single categories. For example, "the final breakthrough in cerebral development" (*1*), "human intelligence" (*8*), and "civilization" (*6*) and other similar terms and phrases were identified with one another and combined under the category "*Homo sapiens*." Next, to avoid redundancy, steps that included other steps were similarly combined. For example, the origin of mitochondria (*7*) and sexual reproduction (*6*, *18*) were categorized under "eukaryogenesis" (*14*) – defined as the evolutionary emergence of the last eukaryotic common ancestor (LECA) from its likely bacterial and archaeal ancestors (*196*) – since LECA already possessed mitochondria (*209*) and reproduced sexually (*210*, *211*), while its free-living bacterial and archaeal ancestors did not. As a result of these combinations, the total number of hard steps (*n*) proposed by certain references were reduced, with both the revised and original *n* estimates displayed at the bottom. The candidate hard steps are listed in rough chronological order, going from top to bottom.

| **Candidate Hard Step** | **Carter 1983 (*1*)** | **Barrow & Tipler 1986 (*7*)** | **McKay 1996 (*8*)** | **Carter 2008 (*6*)** | **McCabe & Lucas 2010 (*13*)** | **Lenton & Watson 2011 (*14*)** | **Lingam & Loeb 2019 (*17*)** | **Snyder-Beattie *et al.*, 2021 (*18*)** | **Sum** |
|---|---|---|---|---|---|---|---|---|---|
| Abiogenesis† | X | X | X | X | X | X | X | X | 8 |
| Glucose fermentation to pyruvic acid | | X | | | | | | | 1 |
| Oxygenic photosynthesis | | X | X | | X | X | X | | 5 |
| Aerobic respiration | | X | | | | | | | 1 |
| Great Oxidation Event | | | X | | X | | | | 2 |
| Eukaryogenesis† | | X | | X | X | X | X | X | 6 |
| Complex (animal) multicellularity† | | | X | X | X | | X | | 4 |
| Eye precursor | | X | | | | | | | 1 |
| Chordates | | X | | | | | | | 1 |
| Endoskeleton | | X | | | | | | | 1 |
| Land ecosystems | | | X | | | | | | 1 |
| Animal intelligence | | | X | | | | | | 1 |
| *Homo sapiens* | X | X | X | X | X | X | X | X | 8 |
| **[Revised] *n* =** | **2** | **9** | **7** | **4** | **6** | **4** | **5** | **3** | |
| **[Original] *n* =** | **2** | **10** | **8** | **6** | **10** | **4** | **5** | **4** | |

†Designates a major transition in evolution (Box 3) as defined by Herron 2021 (*49*).

**Table 2. The Major Transitions in Evolution (MTE).** The MTE as defined by Herron 2021 (*49*), showing the overlap with the original list from Maynard Smith & Szathmáry (*61*), as well the updated list from Szathmáry 2015 (*65*).

| MTE (origin of…) | Maynard Smith & Szathmáry 1995 (*61*) | Szathmáry 2015 (*65*) | Herron 2021 (*49*) |
|---|---|---|---|
| Protocells/life† | X | X | X |
| Chromosomes† | X | X | X |
| Eukaryotes† | X | X | X |
| Multicellularity† | X | X | X |
| Eusociality | X | X | X |
| Plastids | | X | X |
| Mutualisms | | | X |
| Colonial animals | | | X |

†Designates major transition that directly led to *Homo sapiens*.